\documentclass[useAMS,usenatbib]{mn2e}
\usepackage{graphicx}
\def\teff{\ifmmode T_{\rm eff} \else $T_{\mathrm{eff}}$\fi}

\def\msun{\ifmmode M_{\odot} \else M$_{\odot}$\fi}
\def\msunyr{\ifmmode M_{\odot} {\rm yr}^{-1} \else M$_{\odot}$ yr$^{-1}$\fi}
\def\zsun{\ifmmode Z_{\odot} \else Z$_{\odot}$\fi}
\def\lsun{\ifmmode L_{\odot} \else L$_{\odot}$\fi}

\def\mup{\ifmmode M_{\rm up} \else M$_{\rm up}$\fi}
\def\mlow{\ifmmode M_{\rm low} \else M$_{\rm low}$\fi}

\newcommand{\egll}{\ifmmode e^{\rm 1G}_{\rm LL} \else $e^{\rm 1G}_{\rm LL}$\fi}
\newcommand{\esll}{\ifmmode e^{\rm 2G}_{\rm LL} \else $e^{\rm 2G}_{\rm LL}$\fi}
\newcommand{\fgll}{\ifmmode f^{\rm 1G}_{\rm LL} \else $f^{\rm 1G}_{\rm LL}$\fi}
\newcommand{\ngll}{\ifmmode n^{\rm 1G}_{\rm LL} \else $n^{\rm 1G}_{\rm LL}$\fi}
\newcommand{\nggll}{\ifmmode n^{\rm 2G}_{\rm LL} \else $n^{\rm 2G}_{\rm LL}$\fi}
\title[A new perspective on globular clusters]{A new perspective on globular clusters, their initial mass function, and their contribution to the stellar halo and to cosmic reionisation}

\author[D.~Schaerer \& C.~Charbonnel]{Daniel Schaerer$^{1,2}$\thanks{daniel.schaerer@unige.ch (DS)}
and
Corinne Charbonnel$^{1,2}$ \\
$^1$Geneva Observatory, University of Geneva,
51, Ch. des Maillettes, CH-1290 Versoix, Switzerland \\
$^2$CNRS, IRAP, 14 Avenue E. Belin, F-31400 Toulouse, France}

\begin{document}

\date{MNRAS, accepted}
\maketitle

\begin{abstract}
We examine various implications from 
a dynamical and chemical model of globular clusters (GCs), which successfully reproduces the observed 
abundance patterns and the multiple populations of stars in these systems assuming chemical enrichment
from fast rotating massive stars.
Using the model of Decressin et al.\ (2007) we determine the ratio between
the observed, present-day mass of globular clusters and their initial stellar
mass as a function of the stellar initial mass function (IMF). We also compute the
mass of low-mass stars ejected, and the amount of hydrogen ionising photons 
emitted by the proto globular clusters.
Typically, we find that the initial masses of GCs must be $\sim$ 8--10 times (or up to
25 times, if second generation stars also escape from GCs)  larger than the present-day stellar mass.
The present-day Galactic GC population must then have contributed to approximately 
 5--8\%  (10--20\%) of the low-mass stars in the Galactic halo.
We also show  that the detection of second generation stars in the Galactic halo, recently announced by different groups,  
provides a new constraint on the GC initial mass function (GCIMF).
These observations appear to rule out a power-law GCIMF, whereas they are compatible with a log-normal one.
Finally, the high initial masses also imply that GCs must have emitted a large amount of ionising photons in the early Universe.
Our results reopen the question on the initial mass function of GCs, and reinforce earlier conclusions 
that old GCs could have represented a significant contribution to reionise the inter-galactic medium at high redshift.
\end{abstract}

\section{Introduction}
\label{s_intro}
Although for a long time thought to be among the most simple stellar
systems, globular clusters (hereafter GCs) have been subject to
intense studies both observationally and through theory and
simulations.  These include for example detailed work on the stellar
content of GCs and on chemical abundances of GC stars,searches for
viable proto-GCs, studies of dynamical effects on massive star cluster
evolution, cosmological simulations of their formation, estimates of
their contribution to cosmic reionisation, and other related Galactic
and extragalactic astrophysical topics \citep[see e.g.\ reviews
  by][]{Gratton04,Brodie06,Piotto2009,Boily10,Elmegreen10iau266}.

Despite these studies many open questions remain, concerning both GCs
as individual objects and as a collective population.  For example,
the origin of Galactic halo stars and the contribution of GCs to this
population, are still unclear \citep[see
  e.g.][]{Hut92,Parmentier07,Bell08,Boley09}.
Similarly, the shape of the globular cluster initial mass function (GCIMF),  the nature of the present-day globular 
cluster mass function, and  the processes and the timescales responsible for transforming the former into the 
latter, are debated \citep[see][]{Fall01,Vesperini03,Parmentier05,Parmentier07,Elmegreen10}.
Also, first steps are being made in order to understand GC formation in
cosmological simulations \citep{Bromm02,Kravtsov05,Boley09,Griffen10}.
Finally \citet{Ricotti02}
has shown that GCs emit enough ionising photons to reionise 
the Universe, provided their escape fraction $f_{\rm esc}$ is of order unity. 
Examining this question is also of interest in the present context, where it
appears that galaxies found so far in deep surveys are insufficient to reionise 
the inter-galactic medium \citep[see e.g.][]{Bunker09,Ouchi09,McLure10,Labbe10}, and 
where the main sources responsible of cosmic reionisation, presumably faint, low-mass galaxies,
below the current detection limits \citep[cf.][]{Choudhury07,Choudhury08},
remain thus to be identified.

A major paradigm-shift has occurred recently in the GC community, that
sheds new light on these key questions.  Indeed, detailed abundance
studies of their long-lived low-mass stars made possible with 8-10m
class telescopes, together with high-precision photometry of Galactic
GCs performed with HST, have revolutionised our picture of these
stellar systems.  It is now clear that individual GCs host multiple
stellar populations, as shown by their different chemical properties
and by multimodal sequences in the colour-magnitude diagrams
\citep{Bedin04,Piottoetal07,Milone08,Milone10,VillanovaPiotto10}.
Indeed, although nearly all GCs\footnote{With the notable exception of
  $\omega$ Cen, M22, and M54 \citep[see e.g.][ and references
    therein]{DaCosta_M22_09,JohnsonPilachowski10,Siegel07,Carretta10M54}.}
appear to be fairly homogeneous in heavy elements \citep[i.e.,
  Fe-peak, neutron-capture, and alpha-elements, see
  e.g.][]{James04,Sneden05,Carrettaetal09Fe}, they all exhibit large
star-to-star abundance variations for light elements from C to Al that
are the signatures of hydrogen-burning at high temperature implanted
at birth in their long-lived low-mass stars \citep[see e.g.][ and
  reference
  therein]{Gratton01,Gratton04,Sneden05,PrantzosCCCI07,Carretta09,Charbonnel10}.
In fact, the so-called O--Na anticorrelation is ubiquitous in Galactic
GCs, and is now accepted as the decisive observational criterion
distinguishing {\it bona fide} GCs from other clusters
\citep{Carretta10}.

The current explanation for these chemical patterns is the so-called ``self-enrichment'' scenario that 
calls for the formation of at least two stellar generations in all GCs during their infancy:
The first generation stars were born with the proto-cluster original composition, which is that of contemporary field halo stars, while the second generation stars formed from original gas polluted to various degrees by hydrogen-burning processed material ejected by more massive,
short-lived, first generation GC stars. Details and references 
can be found e.g. in \citet{PrantzosCC06}, who discuss the pros and cons of two 
versions of this ``self-enrichment scenario", invoking either massive AGB stars \citep[e.g.][]{CottrellDaCosta1981,VenturaDAntona2001,VenturaDAntona09}, 
or fast rotating massive stars \citep[e.g.][]{Decressin07b} as polluters.

Whatever the actual polluting stars, an immediate consequence of this scenario is that, in order to reproduce the present proportion of first to second generation stars with acceptable values of the polluters IMF, the initial stellar masses of GCs must have been considerably larger than their present-day value \citep{PrantzosCC06,Decressin07b,Dercole08,Decressin10,Carretta10}. 
However, most of the extra-galactic studies have not yet incorporated this revised picture,
or have not yet explored the resulting implications.
Furthermore, the recent discovery of stars with signatures characteristic of 2$^{\rm nd}$ generation GC stars among the
metal-poor halo population \citep{Carretta10,Martell10} sheds new light on the amount of low-mass stars
ejected from GCs and on the initial mass function of these clusters, as we shall show below.

In the present paper, we explore several consequences of this new paradigm, based on the model
that was developed by \citet[][ hereafter DCM07]{Decressin07b} to describe the early chemical and dynamical evolution of GCs. In this model fast-rotating massive ($M \ga$ 25 \msun) stars are  responsible for the GC pollution.
The model successfully explains the observed abundance patterns of present-day GC stars, and has also been tested with N-body and 
hydrodynamical simulations \citep[see][]{Decressin08,Decressin10}. 
Its main assumptions are briefly described and summarised in \S \ref{s_model}. 
Within this framework, we constrain the relation between the initial and the present stellar mass of GCs (\S \ref{initialmass}), as well as the contribution to the stellar halo (\S \ref{contribhalo}), taking the recent observational identification of second generation stars in the Galactic halo \citep{Martell10, Carretta10}  into account.  
Implications on the GCIMF are derived in \S \ref{GCIMF}. 
Finally, we derive in \S \ref{GCreionisation} a well-defined ionising photon production rate for proto-GCs,
taking all the detailed observational constraints from nearby GCs
into account, and estimate their contribution to cosmic reionisation.
Our main conclusions are summarised in \S \ref{s_conclude}.

\section{The adopted chemical and dynamical evolution model}
\label{s_model}

DCM07 and \citet{Decressin07a} have shown that the O-Na anticorrelation observed in GC stars can be explained if a second generation of low-mass stars 
form from the ejecta of first stellar generation fast rotating massive stars mixed with some original interstellar material. In their model, the first generation forms the full mass spectrum of stars
described by a power-law IMF at the high end and a log-normal below 0.8 \msun.
The second generation of ``polluted'' stars is assumed to form only low-mass stars,
following the same log-normal IMF (see below). 
The model allows for dynamical cluster evolution,
and more specifically for the evaporation of stars due to primordial gas expulsion driven by supernovae as well as for long-term dynamical processes as described by \citet{Decressin08,Decressin10}. 

The main free parameter of the model is the IMF slope $x$ above 0.8 \msun   of the first stellar generation, the low-mass IMF being set for both the first and second stellar populations to the present-day mass function observed in GCs \citep{Paresce00}.
Second generation low-mass stars are then formed from the mass of slow wind ejecta $f_{\rm SW}$ predicted by the 
stellar evolution models of \citet{Decressin07a} and after dilution of this material with interstellar gas.
The parameter $d$ describing this dilution is inferred from the observed Li--Na anticorrelation \citep[see][]{Decressin07b,Charbonnel10b}.
We adopt $d=1.15$ from DCM07 as our standard value and comment on the (relatively weak) dependence of 
our results on this parameter.

Allowing for the escape of a fraction of first generation stars, the model then predicts
the relative number of 1$^{\rm st}$ and 2$^{\rm nd}$ generation stars, as well as detailed abundances ratios 
of these stars, which successfully reproduce observed abundance patterns and anticorrelations
(see DCM07).

The fraction of ``unpolluted'', pristine 1$^{\rm st}$ generation long-lived stars $f_p$ still present today 
in GCs can be determined observationally from the distribution of stars along the O-Na anticorrelation \citep[see e.g.][]{PrantzosCC06}. 
In the nomenclature of DCM07 one has $f_p=\ngll /(\ngll +\nggll )$, where $\ngll$ and $\nggll$
are the number of 1$^{\rm st}$ and 2$^{\rm nd}$ generation low-mass, long-lived stars, respectively.
Observations of the O-Na anticorrelation in a large GC sample by \citet{Carretta10} provide a median value ($\pm$ 68 \% CL) of $f_p=0.33 ^{+0.07}_{-0.08}$ both for the total sample and for the lowest metallicity ([Fe/H] $< -1$) hence oldest subsample.
The semianalytical model of DCM07 predicts $f_p$ as a function
of the IMF slope, the dilution parameter $d$, and the fraction \egll\ of low-mass, 1$^{\rm st}$ generation stars
being lost from the cluster due to dynamical processes (see Eqs.\ 20, 23 of DCM07).
We can therefore invert this problem to determine, for each value of the IMF slope,
the lost stellar mass fraction \egll\ from the observed value of $f_p$.
With this at hand, all the properties of the two stellar generations mixed within the 
GC can be determined (see DCM07). Here we are in particular interested in the relation
between the present-day, observed stellar mass and its total, initial value, as well
as in the mass of stars ejected from the cluster. These are derived below.

To do so, we generalise the dynamical evolution scenario discussed in depth by DCM07 
that allows for mass loss from the cluster (as described by $\egll >0$ in their ``Scenario II")
due to mass segregation and evaporation of stars.
We consider the IMF slope $x$ above 0.8 \msun\ as a free parameter, 
and we determine the allowed values of $x$ from the observed value of 
$f_p$ given above. However, it is understood that most observations 
for proto-GCs indicate an IMF slope close to Salpeter ($x=1.35$) in this 
mass range \citep[see e.g.][]{Chabrier03,DeMarchi10,Bastian10}.
While DCM07 assume that all second generation stars are retained 
within the cluster ($\esll=0$ in their notation), we will
subsequently relax this assumption, motivated by recent findings of
some chemically polluted, second generation stars in the Galactic halo
\citep{Carretta10,Martell10}.


\begin{figure*}
\centerline {\includegraphics[width=8.8cm]{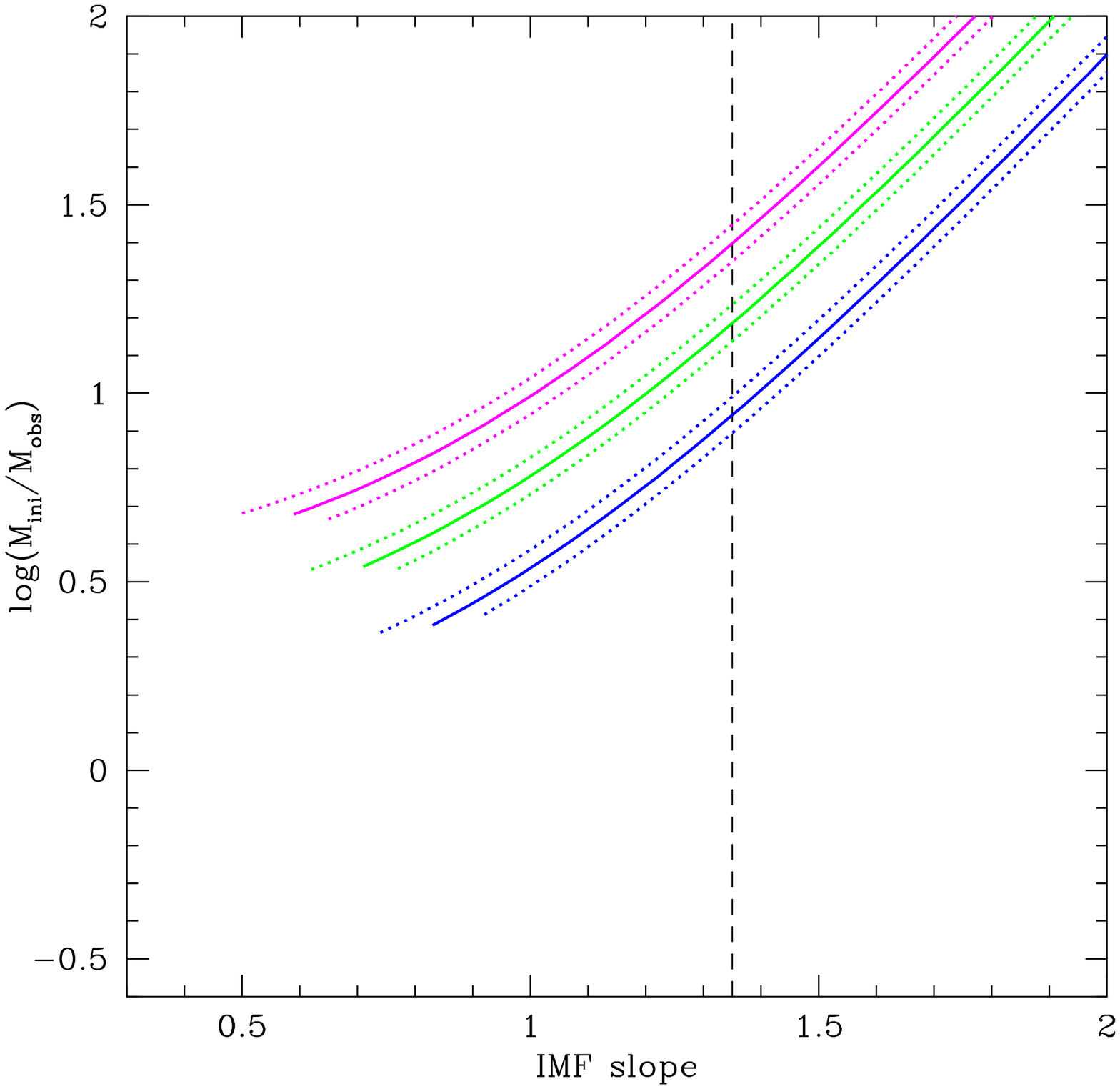}
\includegraphics[width=8.8cm]{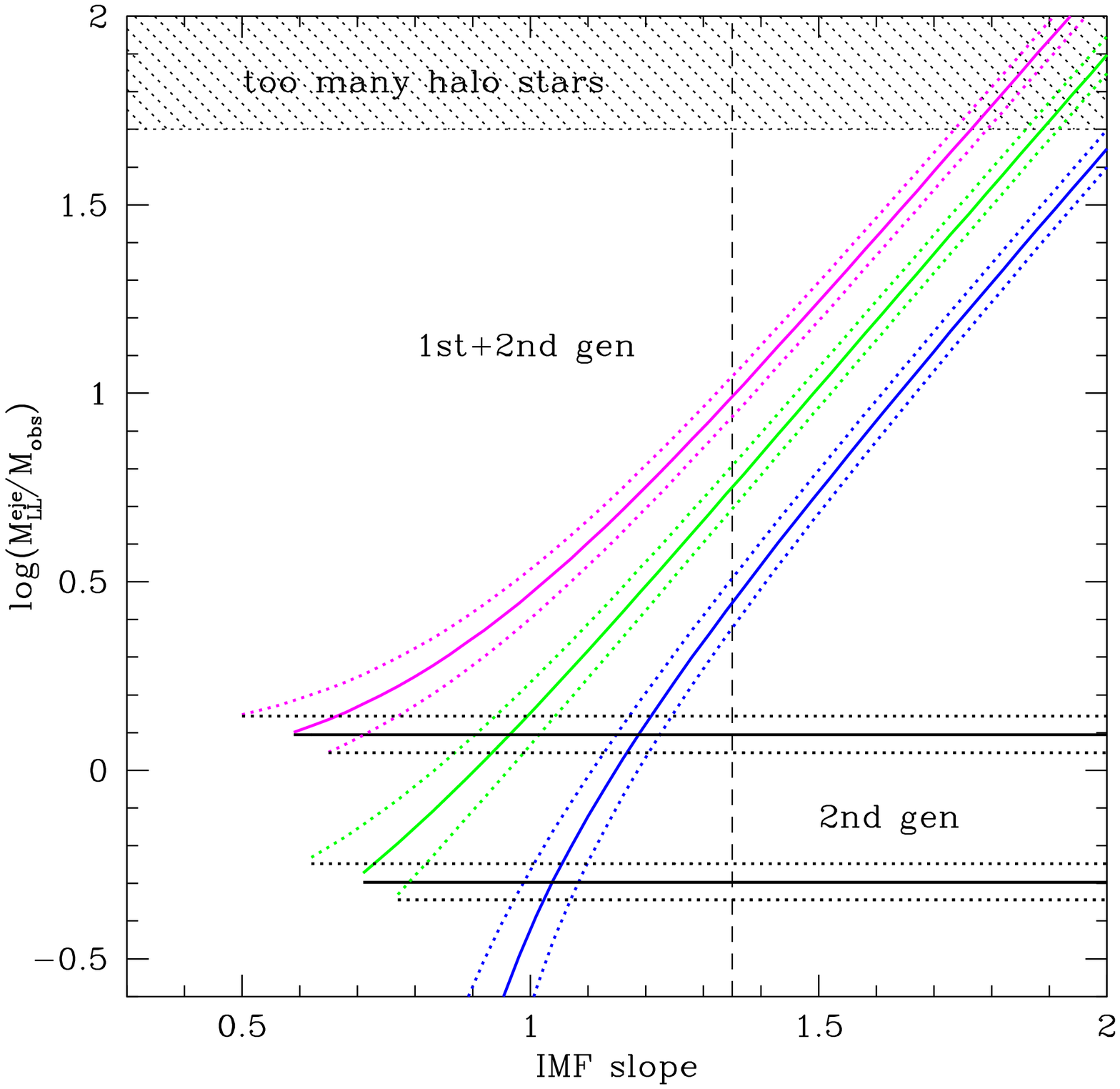}}
\caption{{\em Left:} Ratio between the initial and present-day mass of globular clusters
as a function of the massive star IMF slope (1.35 is the Salpeter value) and for $f_p=0.33 ^{+0.07}_{-0.08}$
(boundaries shown by dotted lines).
The blue lines correspond to the case where no second generation stars escape from the cluster ($\esll=0$), while green and magenta lines show the values predicted when allowing
also for the evaporation of 2nd generation stars with fractions
$\esll=0.43$ and 0.65 respectively.
{\em Right:} Mass ratio between the ejected low-mass stars
(total including 1$^{\rm st}$ and 2$^{\rm nd}$ generation stars) and the present-day mass 
as a function of the massive stars IMF slope.
As for the left figure, blue, green, and magenta lines shown the 
predictions for $\esll=0$, 0.43, and 0.65 respectively, and for the 
observed range of $f_p$.
The black lines show the contribution of 2$^{\rm nd}$ generation stars for the
latter two cases.
The shaded area indicates the region where the present-day GCs would overpredict
the amount of halo stars (assuming 2\% of the stellar halo mass in GCs).
}\label{fig_m}
\end{figure*}

We now follow the semianalytical model of DCM07 and their notation. 
The current mass of 1$^{\rm st}$ generation long-lived (i.e.\ low-mass) stars,
$M_{LL}^{1G}$, in a GC is $M_{LL}^{1G}= f_p \times M_{\rm obs}$, where
$f_p$ is the fraction of 1$^{\rm st}$ generation stars determined
from observations (cf.\ above), and
$M_{\rm obs}$ is the ``observed'', current stellar mass of the globular cluster,
excluding stellar remnants\footnote{Following standard stellar evolution, stellar remnants 
constitute approximately 30\% of the total cluster mass after 12 Gyr. This fraction may, 
however, depend also on the dynamical evolution of the cluster \citep[cf.]{Kruijssen08}.}.
Allowing for dynamical mass loss of stars from the cluster, 
the  total (initial) stellar mass of the cluster can be written as
\begin{equation}
M_{\rm ini} = \frac{f_p}{(1-\egll) \, \fgll} \times M_{\rm obs},
\label{eq_m}
\end{equation}
where \fgll\ stands for the fraction of stellar mass forming
low-mass stars (in the 1$^{\rm st}$ generation), i.e.\ the mass fraction
of the IMF found at $\la 0.8$ \msun, and where \egll\ 
is the fraction of low-mass stars from the 1$^{\rm st}$ generation having
escaped from the cluster during its history.
The total mass of ejected low-mass stars is then
\begin{equation}
M^{\rm LL}_{\rm eje} = \left( \frac{f_p \, \egll}{1-\egll} + \frac{(1-f_p) \, \esll}{1-\esll} \right) \times M_{\rm obs},
\label{eq_meje}
\end{equation}
where we also allow for a fraction \esll\ of low-mass stars of the second
generation to escape (see \S \ref{contribhalo}).

In the dynamical GC scenario discussed by DCM07, 
\egll\ can be determined from the observed fraction $f_p$ of 1$^{\rm st}$ generation stars for a given
slope $x$ of the massive stars IMF, assuming a value for the global dilution factor $d$ (see their Fig.\ 4)\footnote{One obtains 
$\egll = 1- \frac{1}{\fgll}\left(f_{\rm SW}(1+d)(1-\esll)\left[\frac{1}{1-f_p}-1\right]\right)$
from their Eqs.\ 3 and 23, for the assumptions of scenario II, but relaxing the 
hypothesis of $\esll=0$, i.e.\ allowing also for loss of 2$^{\rm nd}$ generation stars.}.
Since \fgll\ depends only on the IMF, it is straightforward to compute 
the relation between the observed and the total initial mass of GCs (Eq.\ \ref{eq_m})
and the amount of stellar mass ejected  (Eq.\ \ref{eq_meje}).

\section{The initial mass of globular clusters}
\label{initialmass}
The ratio between the initial and present-day stellar mass computed in this manner is shown in
the left panel of Fig.\ \ref{fig_m}. 
Blue lines show the case of $\esll=0$ as in DCM07, the green and magenta curves when accounting for $\esll=0.43$ and $0.65$ 
respectively (see below).
As can be seen, the steeper the IMF, the higher the ratio $m$ between the initial and current mass. This is the case since in the present framework massive stars are responsible for the chemical pollution of the GC that leads to the formation of second generation low-mass stars.
Hence for a steeper IMF, a larger total mass of stars is required to compensate the relative decrease of massive stellar polluters in order to reproduce $f_p$. In this way, the same total mass of massive star ejecta incorporated into the second generation stars can be produced.
As mentioned above, all results shown here are computed adopting $d=1.15$ for the dilution parameter.
A stronger dilution of the ejecta from the rapidly rotating massive stars (i.e.\ increasing $d$) 
would imply lower values of $M_{\rm ini}/M_{\rm obs}$ and  $M^{\rm LL}_{\rm eje} /M_{\rm obs}$, since more material is then
available to form the 2$^{\rm nd}$ generation stars. In practice, changing $d$ by a factor 2 (3) around our ``standard'' value implies 
changes of $\sim$ 40--50 (100) \% in the initial and ejected masses, comparable to the differences between
the different cases illustrated in Fig.\ \ref{fig_m}.

For a Salpeter IMF as assumed in DCM07,  the initial cluster mass is found to be $\sim$ 8--10 times larger than the current (observed) mass when no second generation stars are lost, as seen in  Fig.\ \ref{fig_m}.
Considering massive AGBs as potential GC polluters (instead of massive stars), and assuming that none of the second generation stars is lost, \citet{PrantzosCC06} and \citet{Carretta10} also found that the original cluster population should have been larger than the current one by $\sim$ one order of magnitude for a Salpeter IMF.
This agreement is dictated by the amount of initial mass of polluters needed to provide enough material for the second stellar generation.
If we attribute the recently observed 2$^{\rm nd}$ generation stars in the halo to the present population of GCs (cf.\ \S \ref{contribhalo}.)
we must allow for a loss of 2$^{\rm nd}$ generation stars (i.e.\ $\esll>0$), which implies even higher initial masses, as
shown by the green and magenta curves in Fig.\ \ref{fig_m}.
In this case we typically find initial cluster masses (in stars) 15--25 times the present-day mass, for a Salpeter slope.
Of course, the masses of proto-GC clouds must be even higher, depending on their star-formation efficiency.

\section{Contribution to the Galactic stellar halo}
\label{contribhalo}
The ratio between the ejected stellar mass and the present-day mass and its dependence 
on the IMF slope is illustrated in the right panel of Fig.\ \ref{fig_m}.
The total amount of low-mass ($<0.8$ \msun), long-lived stars (both the first and second generation) ejected from the cluster 
shown by the blue lines
corresponds to $\sim (2.5-3.2)$ times the observed, 
present-day mass of globular clusters, for a Salpeter IMF in the case where no second generation stars are lost as assumed by DCM07. 
This amount increases for a steeper IMF.  These stars must contribute to the population of the Galactic halo.

Recent observations have found indications for chemically
polluted, 2$^{\rm nd}$ generation stars in the Galactic halo, with a
frequency of $f_s\sim$ 1.4--2.5\% \citep{Carretta10,Martell10}.
While \citet{Vesperini10} have examined the ejection of these stars
with hydrodynamic cluster models, we follow here a different approach.
If these stars originate from the population of present day GCs,
we can easily {\em infer} the escape fraction of 2$^{\rm nd}$ generation
stars $\esll=(1+(1-f_p)/(f_s \times M_{\rm halo}/M_{\rm now}^{\rm GC}))^{-1}$,
where $M^{\rm now}_{\rm GC}/M_{\rm halo}\approx$ 2\% is the fraction
of present-day GC of the total stellar halo mass
\citep{Freeman02}.
For $f_s=$ 1 (2.5)\% we obtain $\esll=$ 0.43 (0.65), i.e.\
a loss of approximately half of the second generation low-mass stars.
With such a loss, the initial cluster masses and 
the total amount of ejected stars must be even higher than discussed 
above. The corresponding values are shown in Fig.\ \ref{fig_m} with
green and magenta lines.
Typically both initial and ejected mass are increased by a factor
1.7--3.5 compared to the case of \esll=0.
For a Salpeter slope the mass of ejected low-mass stars
is then $\sim$ 5--10 times the present day GC mass.

From the current total mass of halo Galactic globular clusters
of $M_{\rm now}^{\rm GC} \sim 2 \times 10^7$ \msun\ 
and the total halo mass $M_{\rm halo} \approx 10^9$ \msun\
\citep{Freeman02}, i.e.\ the above 2\%, we therefore 
find that low-mass stars ejected from the present-day population 
of GCs make up $\sim$ 5--8\% of the mass of halo stars  if \esll=0, 
or 10--20\% (for the above values $f_s$) if all halo 
2$^{\rm nd}$ generation stars also come from these clusters.
These numbers could be a factor 0.75 lower if a lower mass-to-light ratio
of $M/L_V=1.5$ \citep[cf.]{Dubath97,Larsen02} instead of 2 adopted by
\citep[][and others]{Freeman02} was more appropriate.

For comparison, \citet{Carretta10} estimate a minimum contribution
of GC stars to the stellar halo of 2.8\%, but up to a factor $\sim$ 10 more,
while from \citet{Martell10} one obtains a contribution of $\sim$ 17.5\%
\footnote{Their estimated 50\% of the halo mass corresponds to the
{\em total} stellar mass including the full mass spectrum. For a Salpeter slope
one has a fraction $\fgll \sim$ 0.35 of low-mass stars; i.e.\ 17.5\%.}.
These estimates, based on the observed fraction of 2$^{\rm nd}$ generation stars
in the halo and on models accounting for multiple stellar generations, agree with ours.
Other authors, using calculations of the GC survival fraction or their
destruction rate \citep[cf.][]{Gnedin97}, estimate contributions of $\la 10$\% \citep{Parmentier07} 
or less \citep[3-8 \%]{Boley09} to the Galactic halo from the present-day GC population.
Other estimates, based on a variety of different initial mass functions of
stellar clusters range e.g.\ from 4--40 \% \citep{Fall01}, $\sim$ 40--50\% 
\citep{Baumgardt08},  or up to 30--80 \%
 \citep{Boley09}. However, these studies neglect  the observed 
multiple stellar generations in GC and their implications.

\section{Implications for the initial globular cluster mass function}
\label{GCIMF}

The fact that GCs show -- now basically by definition \citep[cf.][]{Carretta10} -- stellar 
generations, distinct by their chemical abundances, allows us to take a new step
in constraining their initial cluster mass function.
Indeed since stars showing the abundances characteristic of 2$^{\rm nd}$ generation stars
have recently been found in the Galactic halo \citep{Carretta10,Martell10}, their
frequency among normal halo stars provides interesting, new constraints
within the framework of the model examined here.

Consider two limiting cases. First let us assume that {\em all} 2$^{\rm nd}$ generation stars found
in the halo originate from the present-day population of GCs. As discussed above,
one then finds for the initial mass of GCs typically $M_{\rm ini} \approx (15-25) \times M_{\rm obs}$.
The globular cluster initial mass function (GCIMF) must then be equal to the observed one
-- commonly described by a log-normal with a characteristic mass $M_c \sim 1.4 \times 10^5$ 
\msun\ \citep[cf.][]{Harris91} --
but with $M_c$ increased by a factor 15--20, i.e.\ $M_c \sim (2.1-3.5) \times 10^6$ \msun.
Besides this, there is no room left for other proto-GCs, since these  would otherwise 
contribute -- by definition -- additional 2$^{\rm nd}$ generation stars to the halo.

Now assume the other extreme, namely {\em none} of the observed 2$^{\rm nd}$ generation halo stars
are from the observed GCs.  
This corresponds to \esll=0, and
we know then that the present-day GC population had an initial mass function
with a characteristic mass $\sim$ (8--10) times the present value
($M_c \sim (1.1-1.4) \times 10^6$ \msun) and a total mass of 
$M_{\rm ini}^{GC} \sim (1.6-2) \times 10^8$ \msun.
In addition, however, other, dissolved GCs must be invoked to explain the presence of 
these peculiar stars in the halo. Let us assume that the total GCIMF is given by a power law
with a slope $\beta=-2$, as often studied  \citep[e.g.][]{Fall01,Boley09}.
The lowest normalisation we can chose is the one tangential to, i.e.\ osculating the log-normal
initial mass function of the present-day GCs, as discussed e.g.\ in \citet{Boley09}.
From this we compute the total initial mass of GCs to be dissolved, 
$M_{\rm ini}^{\rm GC-diss}$, by subtracting $M_{\rm ini}^{GC}$ from 
$M_{\rm ini}^{\rm tot}$ given by the integral of the GCIMF over the same
range considered by \citet{Boley09}\footnote{Approximately over
$\log(M) \in [3.27,7.27]$, corresponding to a range of magnitudes V from 
-12 to -2 for the present-day GC mass function.}.
Since our osculating mass is (8--10) times higher than the one of \citet{Boley09}
we obtain $M_{\rm ini}^{\rm tot}=(1.6-2) \times 10^9$ \msun, from which 90\% (or more)
is in globular clusters which must be dissolved.  Since the fraction of low-mass
($\la 0.8$ \msun) stars is $\fgll \sim 0.35$ for a Salpeter slope, all the globulars 
then contribute approximately 50--70\% of the present-day stellar halo 
mass\footnote{Assuming $M/L_V=1.5$ instead of 2 (cf.\ above), this percentage would be 
lower by a factor 0.75.}

From Eq.\ \ref{eq_m} and counting the fraction $(1-f_p)$ low-mass stars, we finally obtain the 
total amount of 2$^{\rm nd}$ generation stars produced in these clusters 
$M_{2G}^{\rm GC-diss} \approx 0.075 \times M_{\rm ini}^{\rm tot}$,
which corresponds to a fraction $f_s \sim$ 12--15\%  of the halo mass.
If we adopt a lower mass-to-light ratio ($M/L_V=1.5$) and assume that
30\% of the present GC mass consists of stellar remnants, the expected fraction of 
2$^{\rm nd}$ generation stars in the halo may be somewhat lower, $\sim$ 6--8 \%.
To compute this value we have implicitly assumed -- in the absence of other  information
-- that all GCs can be described by the same values of $f_p$ and \egll\
as those derived from the present-day GCs.

As can be seen, our theoretical prediction for the fraction $f_s$ of 2$^{\rm nd}$ generation stars in the 
halo is considerably larger than the current observational values of $\sim$ 1.4--2.5 \%
from \citet{Carretta10} and \citet{Martell10}. 
The simplest conclusion from this contradiction
is that the GCIMF cannot be a simple power law with $\beta=-2$, as suggested by numerous authors
\citep[cf.][]{Fall01,Boley09,Elmegreen10},
at least not over the mass range considered here. 
However, to reconcile our prediction with the observed value of $f_s$, one would need to
strongly reduce range of the initial cluster masses, since each decade in mass contains
the same amount of total mass for this power law distribution.
In other words, reducing the predicted $f_s$ by a factor 4 or more would imply
an initial cluster mass function over less than 1 dex, compared to our assumption
of $\log(m) \in [3.3,7.3] \msun$.
Similarly, postulating e.g.\ that clusters below a certain mass (say the present day value of
$M_{\rm obs} \la 4. \times 10^4 \msun$ suggested by \citet{Carretta10}) will not become 
globulars, does not solve our problem.
Alternatively, in most clusters the fraction $f_p$ of unpolluted stars could be higher than the value 
observed in present-day GCs, in which case one could avoid ``overproducing'' the number of 
2$^{\rm nd}$ generation stars in the halo.
In this case, however, we cannot properly speak of an initial mass function for GCs,
since these objects with much higher values of $f_p$ cannot be the progenitors of the
present-day GC population.

We are therefore naturally drawn to abandon the picture of a ``universal'' initial power law mass function for
all clusters, including super star clusters, young massive clusters etc.\ and for progenitors of 
present-day GCs. 
Then, as already discussed above, the observations of the 2$^{\rm nd}$ generation halo stars can be 
understood if the GCIMF is log-normal, as e.g.\ proposed by \citet{Vesperini03} and 
\citet{Parmentier05,Parmentier07}.
However, other initial mass functions, e.g.\ a power-law with a turn-over or
Schechter-type functions, cannot yet be excluded.

In any case we have shown here that the observed fraction 2$^{\rm nd}$ generation stars 
in the halo can in principle provide very useful information on the distribution of the
initial masses of globular clusters, the GCIMF.
Of course, our analysis does not constrain the initial mass function of other
(non-globular) clusters. In fact, since the percentage of halo stars
originating from GCs is typically $\la$ 20\%  in our scenarii, there is room
for other clusters, accreted satellites, or others to provide the rest
of the present-day stellar halo.
After many recent studies proposing a ``unified'' picture for the formation
and evolution of clusters of all kinds including GCs 
\citep[cf.][and references therein]{Brodie06,Meurer95,Fall01,Vesperini03,Elmegreen10}, 
it may well be that recent progress on GC stars and the finding of 2$^{\rm nd}$ generation stars among 
the halo population forces us again to revise this picture.
One of the main questions arising now is actually what distinguishes 
``normal'' clusters from globulars and ``globulars-to-become'', i.e.\ what causes a cluster 
to form one or two separate stellar populations. Is this e.g.\ related to their initial central density,
to external conditions, or maybe to completely different formation scenarios,
as e.g.\ suggested by \citet{Searle78,Freeman93,Boeker08}  or can this be understood
within the framework of current hydrodynamic and cosmological formation
models \citep[e.g][]{Boley09,Elmegreen10}?

New detailed (hydro-)dynamic models of cluster formation and evolution
taking into account recent insights gained from scenarii explaining 
the detailed behaviour of observed abundance pattern in GC stars
\citep[e.g.][]{Decressin08,Dercole08,Decressin10} are clearly likely needed to
progress further on this issue. In parallel it will be useful to firm up the 
first studies of second generation stars found in the Galactic halo,
as they currently suffer e.g.\ from poor statistics \citep{Carretta10}
or from uncertainties in observational criteria identifying these stars
\citep{Martell10}.

\begin{figure}
\includegraphics[width=8.8cm]{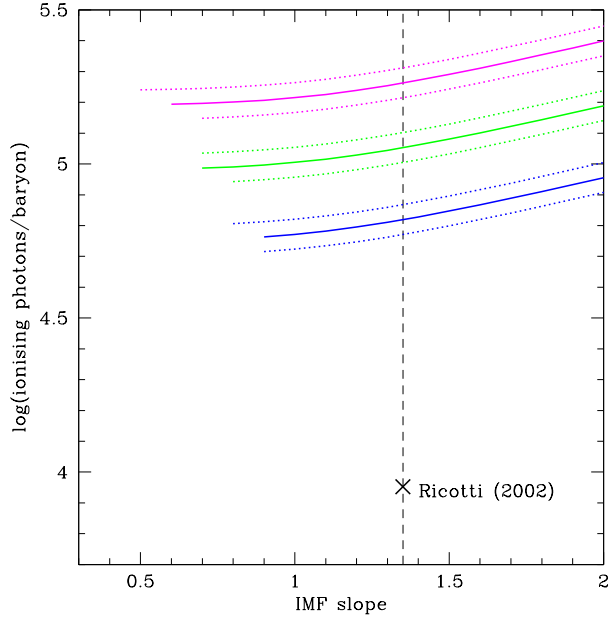}
\caption{Ionising photon production of globular clusters normalised
per baryon currently locked up in stars, $\log \eta^\prime$, plotted as function
of the massive stars IMF slope $x$. 
Blue, green, and magenta lines show respectively values of 
$\eta^\prime$ for IMF slopes allowing one to reproduce the observed fraction 
$f_p$ (boundaries shown by dotted lines) assuming $\esll=0$, 0.43, and 0.65 respectively.
The value of $\eta$ from \citet{Ricotti02} computed for metallicity $Z=0.03$ \zsun\ and the Salpeter slope ($x=1.35$) is shown for comparison.}
\label{fig_q0}
\end{figure}

\section{The contribution of globular clusters to reionisation}
\label{GCreionisation}

Since the initial masses of GCs may be substantially larger
than their present day values, their output of ionising radiation
and their contribution to cosmic reionisation also needs to be revised.

In Fig.\ \ref{fig_q0} we plot the predicted H ionising photon output
(i.e.\ the total number of photons emitted above 13.6 eV) during 
the life of a GC normalised to its {\it current} number of baryons\footnote{To convert 
photon/baryon into photon/mass one has e.g.\
$1 \msun=1.19 \times 10^{57}$ baryon.}, $\eta^\prime$, as a function of the
IMF slope for the dynamical scenario described previously.
The Lyman continuum flux was computed using the evolutionary synthesis
code of \cite{Schaerer03,Raiter10} for a low metallicity $Z=1/20 \zsun$ typical
of GCs\footnote{Adopting a different metallicity leads to
small changes (typically $\la$ 0.1--0.2 dex). Similarly, using 
the stellar evolutionary tracks of fast rotating stars used in the
study of DCM07 would lead to relatively small changes compared to 
the effects discussed here.}.

Interestingly, the predicted ionising photon output is quite independent
of the IMF slope (see Fig.\ \ref{fig_q0}), since both the metals 
explaining the observed abundance pattern and the ionising photons
are made primarily in massive stars. 
Should the ``pollution'' of 2$^{\rm nd}$ generation stars be related to massive AGB stars,
we expect a similarly large emission of ionising photons per baryon, since 
the ratio between the initial and present-day mass is similar in both scenarios
(cf.\ above). In the AGB scenario the dependence of $\eta^\prime$ on the IMF
should, however, but somewhat stronger, since the masses responsible
for the stellar ejecta and the ionising photons are more distinct than in the
fast rotating, massive star scenario.
Of course, one also finds an increased output of ionising photons per unit {\em present-day mass} (or baryon number) 
if loss of second generation stars is allowed (green and magenta lines) compared to the case where $\esll=0$ (blue lines). 
This increase is simply due to higher {\em initial} mass of GCs, discussed previously.

\citet{Ricotti02} has estimated the output of GCs
per baryon as $\eta^\prime = \eta \times f_{di}$, with $\eta \approx 9000$\footnote{Making the
same assumptions as Ricotti we confirm this value.}
and $f_{di} \sim$ 2--10 (with a maximum of $f_{di} <100$), where $f_{di}$
-- the equivalent of our ratio $M_{\rm ini}/M_{\rm obs}$ -- is a factor accounting for 
dynamical disruption of GCs during their lifetime.  
In our case, the total photon output is
$\log(\eta^\prime) \approx$ 4.8--4.9 photon/baryon if all second generation
stars remain within the cluster, which is
approximately a factor 5 higher than the typical value adopted 
by \citet{Ricotti02} for $f_{di} \sim 2$.
The amount of emitted ionising photons must be even higher
if 2$^{\rm nd}$ generation stars were lost from these clusters ($\esll >0$), 
as illustrated by the green and magenta lines.

Using  simple but elegant arguments to estimate the number of ionising photons 
emitted per baryon during the formation of GCs, and using a Press-Schechter model 
to compute the GC formation rate, \citet{Ricotti02} has estimated the 
contribution of GCs to cosmic reionisation.
He has shown that the old GCs produced enough ionising photons to reionise the inter-galactic
medium at $z \approx 6$, if the escape fraction of ionising 
photons, $f_{\rm esc}$, from these objects was close to unity.
Our finding of a high ionising photon production in GCs
reinforces the conclusions of \citet{Ricotti02} and leaves 
room for lower values of $f_{\rm esc}<1$ or for other less
favourable assumptions (e.g.\ uncertainties in the age of GCs).
It appears that old GCs, formed as massive super star clusters 
shortly after the Big Bang, could provide a significant if not 
dominant source of UV radiation to reionise the Universe at high redshift,
In any case, it seems unavoidable to seriously consider their
contribution.

With a typical mass of $\sim 1.4 \times 10^6 \msun$ a young 
proto-GC is expected to have a peak UV (say 1500 \AA) magnitude of $M_{\rm AB} \approx -16.7$
at an age $\sim 2-3$ Myr, before fading by $\sim$ 4 mags within $\sim 20$ Myr just 
due to stellar evolution \citep[see e.g.\ models of][]{Starburst99,Schaerer03} or faster when evaporation 
(mass loss) sets in.
At redshifts of $z \sim$ 7--10 this would correspond to a typical UV restframe magnitude
of $m_{\rm AB} \sim 30.3$ at the peak, just slightly fainter than the current detection limits of the deepest
near-IR images taken with the WFC3 camera onboard HST \citep[cf.][]{Bouwens10}.
In any case, single, massive proto-GCs during their youth might be detectable {\em in situ} with
current instrumentation, and are certainly well within the reach of even deeper near-IR observations,
which will be achieved with the JWST.
However, whether the two proposed scenarii (massive AGB or fast rotating, massive stars as the origin
of the bulk of material out which 2$^{\rm nd}$ generation stars form in GCs) can be distinguished
observationally at high redshift, appears {\em a priori} quite difficult, if not impossible.

\section{Summary and conclusions}
\label{s_conclude}
In light of the recently recognised, general existence of multiple stellar generations in globular clusters
(GCs) implying significant losses of 1$^{\rm st}$ generation stars from these clusters, we have re-examined
the initial masses of GCs, the contribution of low-mass stars ejected from GCs to the 
stellar halo of our Galaxy, and the contribution of GCs to the ionising photon production necessary
to reionise the inter-galactic medium at high redshift. 

These quantities have been estimated
from the chemical and dynamical model of \citet{Decressin07b}, which successfully reproduces the 
main observational constraints from 1$^{\rm st}$ and 2$^{\rm nd}$ generation stars, by invoking pollution
from fast rotating massive stars.
The main free parameters of this model are the slope of IMF for high masses ($>$ 0.8 \msun,
the IMF being fixed to the observed log-normal distribution for lower masses), the
relative number of 1$^{\rm st}$/2$^{\rm nd}$ generations stars, given by the fraction $f_p=0.33 ^{+0.07}_{-0.08}$ 
of 1$^{\rm st}$ generation stars determined from the detailed spectroscopic observations of \citet{Carretta10},
and a dilution parameter $d\approx 1.15$  inferred from the Li-Na anticorrelation observed in 
GCs \citep{Decressin07b,Charbonnel10b}.

The dynamical scenario we have explored allows for the evaporation of stars
from the 1$^{\rm st}$ generation (corresponding to an escape fraction of 2$^{\rm nd}$ generation stars of zero, $\esll=0$), 
or from both generations, as suggested by recent observations finding stars characteristic of the 
2$^{\rm nd}$ generation in GCs in the Milky Way halo \citep{Carretta10,Martell10}. The latter case translates to
$\esll \sim$ 0.43--0.65.

We have obtained the following main results for an IMF with a Salpeter slope above 0.8 \msun:
\begin{itemize}
\item The initial stellar masses of GCs must have been $\sim$ 8--10 times larger than the current (observed) mass,
when no second generation stars are lost, in agreement with the earlier results of \citet{PrantzosCC06} and \citet{Carretta10}.
If all 2$^{\rm nd}$ generation halo stars originate from the present population of GCs, the initial cluster masses must have
been $\approx 25$ times larger than the current mass.

\item The mass in low-mass stars ejected from GCs must be $\sim 2.5-3.2$ times their observed, 
stellar mass if all 2$^{\rm nd}$ generation stars were retained, or $\sim$ 5--10 times the present day mass
if $\esll \sim$ 0.43--0.65.
These numbers translate to a contribution of  5--8\%  or 10--20\% respectively of the ejected low-mass
stars to the Galactic stellar halo mass. We have compared our estimate with earlier
values obtained from various methods (cf.\ \S \ref{contribhalo}).

\item The observations of 2$^{\rm nd}$ generation stars in the Galactic halo can constrain the
initial mass function of the GC population (GCIMF). In particular we have shown that 
a power-law with a slope $\beta \approx -2$, as often assumed, is in contradiction with
recent determinations of the fraction of  2$^{\rm nd}$ generation stars in the halo, whereas
a log-normal GCIMF is compatible with these observations.
This finding revives the question about a common mass function and about the physical
processes leading to a distinction between globular clusters with multiple stellar populations
and other clusters.

\item Due to their high initial masses, the amount of Lyman continuum photons emitted
by GCs during their youth must have been substantial. Indeed, we find that their output
corresponds to a total number of ionising photons emitted per baryon, $\eta^\prime 
\approx 10^{4.8-4.9}$ for  $\esll=0$, or $\sim$ 1.7--3.5 times more if $\esll \sim$ 0.4-0.6.
Our results reinforce the conclusion of \citet{Ricotti02} that GCs should contribute 
significantly to reionise the IGM at very high redshift ($z \ga 6$).
Individual, young proto-GCs with typical masses few times $\sim 10^6 \msun$ could just 
be detectable at high redshift in ultra-deep images with the HST, and are certainly within
the reach of the JWST.

\end{itemize}
The dependence of the initial and ejected masses on the  IMF slope has been illustrated in 
Fig.\ \ref{fig_m}. The ionising photon production is found to be quite insensitive to
the high mass IMF, since both the ejecta ``polluting'' the 2$^{\rm nd}$ generation stars and the
Lyman continuum flux originates from massive stars. 
Our main results should also be valid for the massive AGB scenario, at least qualitatively.

\section*{Acknowledgments}
We thank Thibault Decressin, Andrea Ferrara, and Massimo Ricotti 
for comments on an earlier version of this paper.
This work was supported by the Swiss National Science Foundation.

\bibliographystyle{mn2e}
\bibliography{references}

\end{document}